\documentclass[twocolumn,nofootinbib,amsmath,amssymb,a4paper]{revtex4}

\usepackage{graphicx}
\usepackage{dcolumn}
\usepackage{bm}

\def\Bz   {\ensuremath{B^0} }
\def\Bzb   {\ensuremath{\bar{B^0}} }

\def\invfb   {\ensuremath{fb^{-1}} }

\def\invab   {\ensuremath{ab^{-1}} }

\def\bbbar   {\ensuremath{B \bar{B}} }

\def\tbpg   {\ensuremath{2\beta+\gamma} }
\def\dkpi   {\ensuremath{D^\mp K^0 \pi^\pm} }

\def\bdkpi   {\ensuremath{B^0 \to D^\mp K^0 \pi^\pm} }

\def\fis   {\ensuremath{FI} }

\def\mes   {\ensuremath{\mbox{$m_{ES}$}} }
\def\de   {\ensuremath{\mbox{$\Delta E$}} }

\def\md   {\ensuremath{\mbox{$m_{D}$}} }

\def\babar   {\ensuremath{BaBar} }

\begin{document}

\title{\tbpg\ from \bdkpi\ decays at $BaBar$: a simulation study.}

\author{F. Polci, M.H.Schune, A.Stocchi}
 \email{polci@lal.in2p3.fr, schunem@lal.in2p3.fr, stocchi@lal.in2p3.fr}
\affiliation{%
LAL, Laboratoire de l'Accelerateur Lineaire, 94100 Orsay, France}
%
\begin{abstract}
 We present the results of a simulation study to perform the extraction of \tbpg\ from  \bdkpi\ decays through a time-dependent Dalitz analysis  of $BaBar$ data.
\end{abstract}
\maketitle
\section{Introduction}
\begin{figure}
\includegraphics[width=0.4\textwidth]{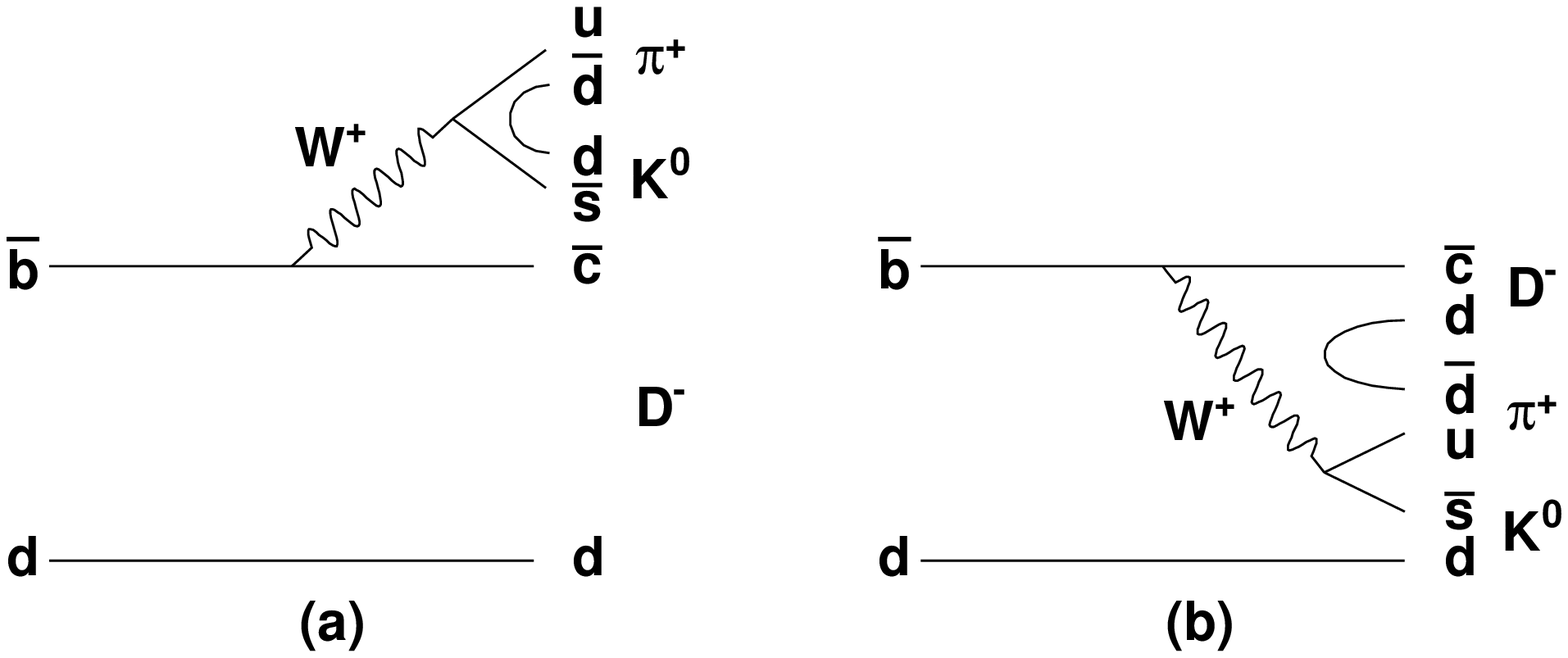}
\includegraphics[width=0.4\textwidth]{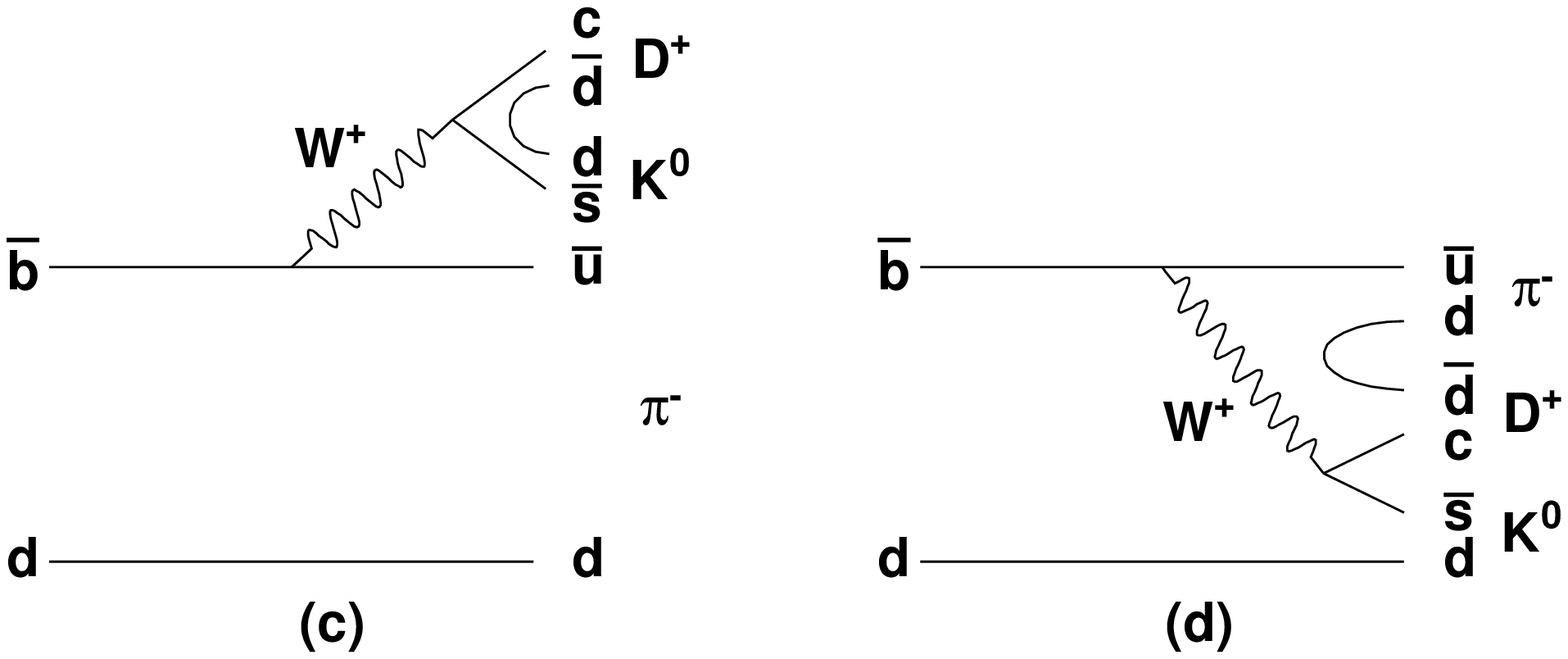}
\caption{\it {Feynman diagrams for the processes contributing to the  \dkpi decays; 
a) $B^0 \rightarrow D^- K^{*+} (K^0 \pi^+)$ and higher $K^{**}$ resonances, 
b) $B^0 \rightarrow \bar{D}^{**0} (D^- \pi^+) K^0$, 
c) $B^0 \rightarrow {D_s}^{**+} (D^+ K^0) \pi^-$, 
d) $B^0 \rightarrow {D}^{**0} (D^+ \pi^-) K^0$. 
The processes in a,b (c,d) are $V_{cb}$ ($V_{ub}$) mediated.}}
\label{DIAGRAMS}
\end{figure}
A decay of a $B^0$ is sensitive to the weak phase $2\beta$ because of the $B^0\bar{B^0}$ mixing. If the decay leads to a final state which can be reached both through  Cabibbo allowed $b \rightarrow c$ ($V_{cb}$) transitions and  Cabibbo suppressed $b \rightarrow u$ ($V_{ub}$) transitions, there is also sensitivity to the weak phase $\gamma$. Thus we have sensitivity to the total phase:
\begin{equation} 
2\beta + \gamma = arg \frac{-V_{td}^{*2} V_{tb}^{2} V_{ub}^{*} V_{us} } { V_{cd}^{*2} V_{cb} V_{cs} } 
\end{equation}
where $V_{i,j}$ are the terms of the CKM quark mixing matrix~\cite{bib:C}~\cite{bib:KM}.
 
Up to now the constraints on $2\beta + \gamma$ come from the analysis of the two-body decays $B^0 \rightarrow D^{(*)} \pi$ and $B^0 \rightarrow D \rho$ \cite{bib:DPi-DRho}. 
Another way of measuring  $2\beta+\gamma$ is to study the \bdkpi decays, as proposed  in \cite{bib:aps-ap}. Figure \ref{DIAGRAMS} shows the Feynman diagrams for the Cabibbo allowed and suppressed processes. Only tree-diagram contributes.
The main advantage of these three-body decays comes from the possibility of performing a Dalitz analysis, which allows to reduce the typical eight-fold ambiguity in the determination of $2\beta + \gamma$  to only a two-fold ambiguity \cite{bib:aps-ap}.
Moreover the ratio $r$ of the Cabibbo allowed over the Cabibbo suppressed amplitudes, which in the two-body decays is about 0.02, is expected here to be quite large (about 0.4) in some regions of the Dalitz plot. This parameter is directly related to the sensitivity to \tbpg.
The ambition of the analysis technique is to be completely model independent, in the sense that at high statistics it will not rely on any theoretical assumption: all quantities will be directly extracted from the fit on data.

A sensitivity study of the channel can be found in~\cite{bib:nousHepPh}.
 Here, following the same approach, we complete it with simulations making use of  realistic background distributions, resolution and tagging performances  taken from the $BaBar$ Run1-4 data sample.
\section {Dalitz contribution to the time dependent equation}\label{sec:timedalitzlik}
The  model assumed for the decay parameterizes the amplitude $A$ at each point $k$ of the 
Dalitz plot as a sum of two-body decay matrix elements and a non-resonant term according
to the isobar model ~\cite{bib:isobar}:
\begin{equation}
A_{c_k(u_k)}e^{i\delta_{c_k(u_k)}}=\sum_j a_je^{i\delta_j} BW^j_k(m,\Gamma,s) + a_{nr} e^{i \delta_{nr}},
\label{eqres}
\end{equation}
where $c_k$ ($u_k$) indicates the Cabibbo allowed (suppressed) decay in each point $k$ of the Dalitz plot. 
Each term of the sum is parameterized with an amplitude ($a_j$ or $a_{nr}$) and a phase ($\delta_j$ or $\delta_{nr}$, where the $nr$ index indicates the non resonant). 
The factor $BW^j_k(m,\Gamma,s)$ 
gives the Lorentz invariant expression for the matrix element of a resonance $j$ as a function of the position $k$
in the $B$ Dalitz plot; the functional dependence varies with the spin $s$ of the resonance, the mass $m$ and the decay width $\Gamma$.

The time dependent evolution  can
 be obtained from the resolution of the
Schroedinger equation, leading to the following expression:
\begin{eqnarray}
P_k(\Delta t, \epsilon,\eta)=\frac{A_{c_k}^2+A_{u_k}^2}{2} &\frac{e^{-\Gamma_B \Delta t}}{4 \tau}
\{1 - \eta \epsilon  C^k \cos(\Delta m_d \Delta t)  \nonumber \\
& +   \epsilon  S^k_{\eta}  \sin (\Delta m_d  \Delta t )\},
\label{timelikeli}
\end{eqnarray}
with :
\begin{eqnarray}
C^k &=& \frac{A_{c_k}^2-A_{u_k}^2}{A_{c_k}^2+A_{u_k}^2}  \nonumber \\
S^k_{\eta} &=& \frac{2 Im (A_{c_k}A_{u_k} e^{i( \tbpg )+\eta i(\delta_{c_k}-\delta_{u_k})})}{A_{c_k}^2
+A_{u_k}^2}
\end{eqnarray}
where $\epsilon=+1$ (-1) if the tagged initial state is a $B^0$ ($\bar{B^0}$), $\eta=+1$ (-1) if the final state contains a $D^+$ ($D^-$),
and $\Delta t$ is the proper time interval  between the reconstructed $B$ ($B_{rec}$) and the tagging
$B$ ($B_{tag}$).
$\Gamma_B$ and $\Delta m_d$ are respectively the $B^0$ decay width and its mixing frequency. 
Thanks to the presence of the terms $BW^j_k(m,\Gamma,s)$ which vary over
the Dalitz plot, we can fit the amplitudes ($a_j$) and the phases ($\delta_j$) of Eq. \ref{eqres}, together
with \tbpg\  with only a two-fold ambiguity.
\begin{table}
\begin{center}
\begin{tabular}{|l|c|c|c|c|c|c|} \hline \hline
                          &$m (\frac{GeV}{c^2})$&$\Gamma (\frac{GeV}{c^2})$&$a(V_{cb})$ &$a(V_{ub})$&$\delta(V_{cb})$ &$\delta(V_{ub})$ \\ \hline
 $D_{s,2^+}(2573)^{\pm}$  &    2.572       &      0.015   &    -     &  0.02    & - & 70   \\ \hline
 $D_{2^+}^*(2460)^{0}$    &    2.461       &      0.046   &   0.12   &  0.048    & 30 & 30   \\ \hline
 $D_{0^+}^*(2308)^{0}$    &    2.308       &      0.276   &   0.12   &  0.048    & 70 &  0    \\ \hline
 $K_{1^-}^*(892)^{\pm}$       &    0.89166     &      0.0508  &     1    &  -        & 0 & -    \\ \hline
 $K_{0^+}^*(1430)^{\pm}$  &    1.412       &      0.294   &    0.6   &  -        & 80 & -   \\ \hline
 $K_{2^+}^*(1430)^{\pm}$  &    1.4256      &      0.0985  &    0.2  &  -         & 0 & -  \\ \hline
 $K_{1^-}^*(1680)^{\pm}$  &    1.717       &      0.322   &    0.3   &  -        & 30 & -   \\ \hline
 Non resonant             &       -        &         -    &    0.07  &  0.028     & 0 & 30  \\ \hline \hline
\end{tabular}
\end{center}
\caption{\it Masses and widths of the resonances considered in the model. 
The last four columns present the chosen values of the coefficients $a_j$ and $\delta$ in 
Eq.~\ref{eqres} for the Cabibbo allowed and Cabibbo suppressed decays respectively.}
\label{tab:resonances}
\end{table}
\section{The \bdkpi Dalitz model}\label{sec:dalitzModel}
The decay mode under study is assumed to proceed through the resonances listed in Table \ref{tab:resonances}.
 The $K^0 \pi^+$ resonances ($K^*$ like) can only come from $V_{cb}$ mediated processes, while both $V_{cb}$ and $V_{ub}$ 
transitions contribute to  $D^- \pi^+$ resonances ($D^{**}$ like). Finally the $D^- K^0$ resonances ($D^{**}_s$-like) come 
only through $V_{ub}$ mediated processes.
Contributions coming from very wide resonances like higher $K$ excited states or 
higher $D^{**}$ excited states will be taken into account by a generic ``non resonant'' term. 

The total phase and amplitude are arbitrary, so we can chose amplitude unity and phase zero for the mode
$K^{*+}(892)$ decaying into $K^0 \pi^+$. All the other amplitudes and phases  values are referred to the $K^{*+}(892)$  ones. 
Since the strong phases are not known experimentally, their values are  
\begin{figure}
\begin{center}
\includegraphics[width=0.4\textwidth]{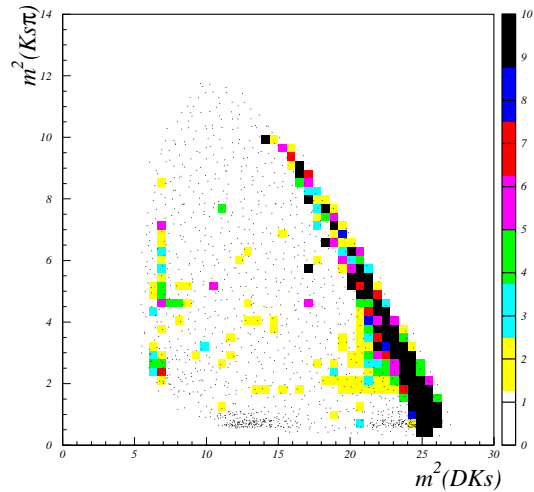}  
\caption{\it  Dalitz distribution of the very high statistics Monte-Carlo
sample of signal events (black dots). Each event is weighted with the second derivative with respect to 2$\beta + \gamma$ of the
log-likelihood (colored squares).}
\label{sensitivity}
\end{center}
\end{figure}
chosen arbitrarily, while the values of the amplitudes comes from the available measurements of related $B$ decay modes and some theoretical considerations~\cite{bib:nousHepPh}. In agreement with the Standard Model we assume for the sensitivity study \tbpg $= 2$ rad.

The total number of \dkpi\ signal events is estimated using the measured 
branching fraction and observed yield~\cite{bib:babarmh}: we expect about 250 signal events per unit of 100 \invfb. 
\section{Sensitivity to \tbpg}\label{sensitivitystudy}
In order to show the regions of the Dalitz plot that mostly contribute to the determination of 2$ \beta + \gamma$, a very high statistics Monte-Carlo sample of signal events has been generated 
according to the nominal model described in Section \ref{sec:dalitzModel}. 
Since the uncertainty on \tbpg\ is:
\begin{equation}
\sigma_{\tbpg} = \sqrt{\frac{1}{\sum{\frac{\partial ^2 ln L}{\partial ^2(\tbpg)}}} } 
\end{equation}
we can weight each event by the quantity $weight$, the second derivative with respect to 2$\beta + \gamma$ of the log-likelihood constructed according to Eq.\ref{timelikeli}:
\begin{equation}
  weight=\frac{\partial ^2 ln L}{\partial^2 (\tbpg)}
\end{equation}
 Note that this likelihood considers only  signal contribution and 
does not take into account tagging and resolution effects.\\
 In Figure \ref{sensitivity}, the weighted distribution expressing the sensitivity (colored squares) is superimposed to the distribution of the Monte Carlo events in the   $m^2_{D^+ K^0}$ versus $m^2_{K^0\pi^-}$  plane (black dots). 
The regions with interference between $B^0 \rightarrow \bar{D}^{**0} K^0$ and $B^0 \rightarrow D^{**0} K^0$ color 
suppressed processes (diagonal side of the Dalitz) show the greater sensitivity to $2\beta + \gamma$.
A particularly sensitive zone is at the intersection between $B^0 \rightarrow D^{-} K^{*+}$  and the color 
suppressed $B^0 \rightarrow D^{**0} K^0$ (right bottom corner of the Dalitz plot). Some sensitivity in the vertical left side of the plot is  
present because of the interference of the $D_{s,2}(2573)^{\pm}$ with the excited $K$ resonances 
($K^{*}_0(1430)^{\pm}, K^{*}_2(1430)^{\pm}, K^{*}(1680)^{\pm}$).

\begin{figure}[!tbp]
\begin{center}
\includegraphics[width=0.4\textwidth]{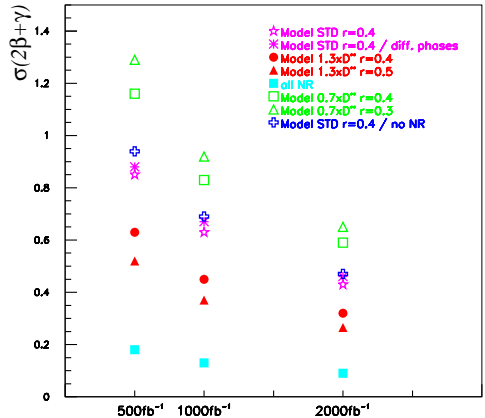}
\end{center}
\caption{\it{  Average absolute uncertainty on $2\beta + \gamma$ as a function of the integrated luminosity. 
Magenta thick stars refers to the model in table \ref{tab:resonances}; thin stars are obtained 
by varying the strong phases values; red full circles and  green open squares come from   
multiplying the $D^{**0}$ amplitudes by a factor 1.3 and 0.7 respectively; red full and green open 
triangles refer to previous models considering $r$=0.5 and $r$=0.3 respectively; blue crosses refers to the model in table \ref{tab:resonances} with no non-resonant component. The cyan full squares  correspond to the model with  only non-resonant amplitudes for all $V_{ub}$ processes presented in \cite{bib:aps-ap}: this is in contradiction with present data \cite{bib:babarmh}.}}
\label{fig:errormodel}
\end{figure}

Since the present measurements (\cite{bib:babarmh},\cite{ref:babarChinois}) are 
not sufficient to fix the Dalitz model, the dependence on the determination of \tbpg\  from the assumed Dalitz structure of the
decay  has been evaluated, has shown on Figure~\ref{fig:errormodel}. Here \tbpg\ is  fit leaving fixed all the other parameters, and 
the impact on the error on 2$\beta + \gamma$ of variations in the decay model is studied as a function of the luminosity.  
\begin{figure}[!tbp]
\begin{center}
\includegraphics[width=0.4\textwidth]{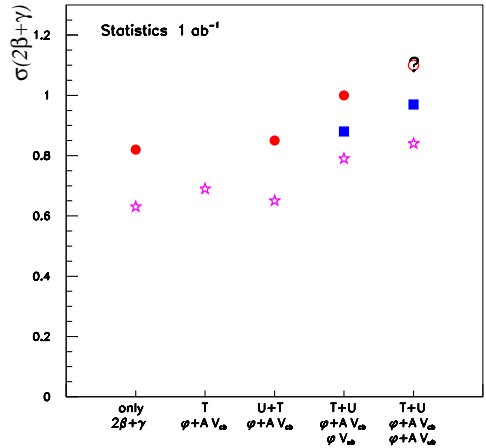}
\end{center}
\caption{ \it{ Average absolute uncertainty on \tbpg\ as a function of the fit configuration (``T'' stands for tagged and ``U'' for untagged events).
The magenta thick stars refers to fit with no background while red dots (blue squares) correspond to fits
where the level of background (flat in the Dalitz plot) is set at 50$\%$ (30$\%$).}}
\label{plot-error}
\end{figure}
 The precision on \tbpg\ strongly depends on the variation of the 
branching fractions of neutral $B$ into $D^{**}$ states within the measured errors. 

 An important feature of this method is the possibility to have a model independent determination of
\tbpg\ by fitting the parameters of the decay model. This is shown on Figure~\ref{plot-error}. 
Considering a sample corresponding to 1 \invab of collected statistics, 
the reference error is obtained as previously by fitting only \tbpg\ with all the
other parameters fixed. Then the other parameters which characterize the decay model are progressively released.
At this point  untagged events are useful: even if they do not carry  information on \tbpg, still they help in determining amplitudes and phases, lowering the uncertainty on \tbpg. 

 A first attempt to estimate the effect of the background is done  generating it flat over the Dalitz plot. 
 The signal over background ratio is fixed to 30$\%$ or 50$\%$. The result, shown in fig.~\ref{plot-error}, is that the error on 
\tbpg\ increases by 25$\%$ and 50$\%$, respectively. 
\section{Simulation}
The previous study has been completed performing new simulations, which take into account realistic background levels and distributions over the Dalitz plot, as well as the resolution effects, as obtained from studies on BaBar data and Monte Carlo samples.

Some variables are useful to discriminate between signal and background: the beam-energy substituted mass $\mes  \equiv \sqrt{(\sqrt{s}/2)^2-{p^*_B}^2}$;  the difference between the $B^0$ candidate's measured
energy and the beam energy $\de \equiv E^*_B - \sqrt{s}/2$ (the asterisk
denotes evaluation in the $\Upsilon(4S)$ CM frame); a fisher discriminant \fis\ combinating some event shape variables; the mass \md of the $D$ meson. 
We define ${\mathcal Y}_{j}^{i}$ as the product
 of the probability density functions (PDFs) of these variables for each 
event $i$ and for each component $j$: signal ($Sig$), continuum background ($Cont$), combinatoric \bbbar decays ($\bbbar$ background) and \bbbar events that peak in \mes\ but not in \de\ signal region (denoted peaking \bbbar background: $Peak$).

The complete expression of the likelihood function, taking into account also the time dependence and the Dalitz distribution is:
\begin{eqnarray}
\ln{ {\mathcal L}} =
 \sum_{i}^{N_c}
 \left[
 \sum_{\Bz\ {\rm tag} }
 { \ln{ {\mathcal L}_{+,i} } }
  +
 \sum_{\Bzb\ {\rm tag} }
 { \ln{ {\mathcal L}_{-,i} } }
 \right]
\label{eq:cp_ll_func}
\end{eqnarray}
where $N_c$ = 7 is the number of tagging categories, ${\mathcal L}_{+,i}$ (${
\mathcal L}_{-,i}$) is the likelihood function for an event in the tagging category $i$ with $B_{tag}
= B^0$ ($B_{tag} = \Bzb$) and:
\begin{eqnarray}
{\mathcal L}_{\pm,i} =  N^i_{Sig} {\mathcal P}^i_{\pm,Sig} {\mathcal Y}^i_{Sig}
+ N^i_{\bbbar}  {\mathcal D}_{\bbbar} {\mathcal T}^i_{\pm,\bbbar} {\mathcal Y}^i_{\bbbar}\\
+ N^i_{Cont}  {\mathcal D}_{Cont} {\mathcal T}^i_{\pm,Cont} {\mathcal Y}^i_{Cont}\\
+ N^i_{Peak}  {\mathcal D}_{Peak} {\mathcal T}^i_{\pm,Peak} {\mathcal Y}^i_{Peak}
\label{eq:likelihoodAll}
\end{eqnarray}
Here  ${\mathcal P}$ is the time-dependent Dalitz PDF for
signal (eq.~\ref{timelikeli}), ${\mathcal D}$ and ${\mathcal T}$ indicate the Dalitz  and the time 
parameterizations for the backgrounds. 

The simulations are performed taking the ${\mathcal D}$, ${\mathcal T}$ and  ${\mathcal Y}$ shapes for the backgrounds from $BaBar$ data and Monte Carlo simulations, while for the signal ${\mathcal P}$ the model previously described is taken into account. About 1000 samples of signal and background are generated and fitted according to these PDFs.

The results show that an integrated luminosity of 354 \invfb (corresponding to  Run1-4 $BaBar$ data) is not enough to extract all amplitudes and phases from the fit. In particular the non resonant component is not correctly determined.

For this reason other  simulations are performed introducing a parameter $r=0.3$, so that for each resonance the relation: $A(V_{ub})=r \times A(V_{cb})$ holds. The choice of the 0.3 value for $r$ is suggested by the result of the $B^0 \to D^0 K^{*0}$ analysis~\cite{bib:babarshahram}, giving a limit $r<0.4$ @ 90\% CL.

The simulation at 354 \invfb in this configuration works if the amplitudes and phases for the $D^{**}_s$ and the non resonant component are fixed to the model value, as well as all the $V_{ub}$ phases. In this case the $V_{cb}$ amplitudes are determined with about 30\% error, and the error on \tbpg\ is of about 75 degrees. Some biases are observed in the amplitudes determination, which are consistently  
\begin{table}
\begin{center}
\begin{tabular}{|l|c|c|c|c|} \hline \hline
                        &$a(V_{cb})$ &$a(V_{ub})$&$\delta(V_{cb})$ &$\delta(V_{ub})$ \\ \hline
 $D_{s,2}(2573)^{\pm}$  &    -              &  0.002 $\pm$ 0.011    & -          & 0 $\pm$ 155   \\ \hline
 $D_{2}^*(2460)^{0}$    &   0.11 $\pm$ 0.01   &  0.046 $\pm$ 0.019    & 27 $\pm$ 7   & 46 $\pm$ 26   \\ \hline
 $D_{0}^*(2308)^{0}$    &   0.13 $\pm$ 0.02   &  0.047 $\pm$ 0.023    & 70 $\pm$ 11  & 38 $\pm$ 31    \\ \hline
 $K^*(892)^{\pm}$       &     1             &  -                  & 0          & -    \\ \hline
 $K_{0}^*(1430)^{\pm}$  &    0.60 $\pm$ 0.02  &  -                  & 81 $\pm$ 2   & -   \\ \hline
 $K_{2}^*(1430)^{\pm}$  &    0.20 $\pm$ 0.03  &  -                  & 0 $\pm$ 2    & -  \\ \hline
 $K^*(1680)^{\pm}$      &    0.30 $\pm$ 0.01  &  -                  & 28 $\pm$ 4   & -   \\ \hline
 Non resonant           &  0.09 $\pm$ 0.02  & 0.070 $\pm$ 0.030     & 357 $\pm$ 13 & 48 $\pm$ 24 \\ \hline \hline
\end{tabular}
\end{center}
\caption{\it Results of the fit simulation corresponding to an integrated luminosity of 10\invab}
\label{tab:resHighStat}
\end{table}
reduced if the simulation is performed at a statistics of 1\invab: in this case the error on \tbpg\ is about 57 degrees.

The last test performed is a simulation at very high statistic (10\invab), where we do not use the parameter $r$, but we fit directly all amplitudes and phases generated at the values presented in table~\ref{tab:resonances}. Results, shown in table~\ref{tab:resHighStat}, show that all the parameters can be fitted, so that \tbpg\ is determined with an error of about 14 degrees in a completely model independent way. Of course the estimation on the error on \tbpg\ depends strongly on the Dalitz model assumed: reality could be different, depending on the relative abundances of the resonances. 

In conclusion, a model independent analysis method for the determination of \tbpg\ has been presented and validated using realistic simulation, based on the observed shapes of the discriminating variables on $BaBar$ data and simulated samples as well as on BaBar tagging and resolution performances. At the luminosity of 1\invab, \tbpg\ can be determined with an error of about 57 degrees.

\end{document}